\documentstyle [11pt,graphicx] {article}
\input{epsf}
\begin{document}
\renewcommand{\baselinestretch}{1.5}
\begin{center}
\huge {\bf Characterizing the non-linear growth of large-scale
structure in the Universe}
\end{center}

\normalsize

\vspace{1cm}

\begin{center}

{\bf Peter Coles,\\ School of Physics \& Astronomy,\\ University
of Nottingham,\\ University Park,\\ Nottingham NG7 2RD,\\ United
Kingdom.\\}

\vspace{1cm}

 {\bf and}\\

\vspace{1cm}

{\bf Lung-Yih Chiang,\\ Astronomy Unit, \\ School of Mathematical
Sciences,\\ Queen Mary \& Westfield College, \\ University of
London, \\ London E1 4NS,\\ United Kingdom.\\}

\end{center}

\vspace{1cm}

\noindent{\bf The local Universe displays a rich hierarchical
pattern of galaxy clusters and superclusters$^{1,2}$. The early
Universe, however, was almost smooth, with only slight 'ripples'
seen in the cosmic microwave background radiation$^{3}$. Models of
the evolution of structure link these observations through the
effect of gravity, because the small initially overdense
fluctuations attract additional mass as the Universe
expands$^{4}$. During the early stages, the ripples evolve
independently, like linear waves on the surface of deep water. As
the structures grow in mass, they interact with other in
non-linear ways, more like waves breaking in shallow water. We
have recently shown$^{5}$ how cosmic structure can be
characterized by phase correlations associated with these
non-linear interactions, but hitherto there was no way to use that
information to reach quantitative insights into the growth of
structures. Here we report a method of revealing phase
information, and quantify how this relates to the formation of a
filaments, sheets and clusters of galaxies by non-linear collapse.
We use a new statistic based on information entropy to separate
linear from non-linear effects and thereby are able to disentangle
those aspects of galaxy clustering that arise from initial
conditions (the ripples) from the subsequent dynamical evolution.}

In most popular versions of the ``gravitational instability''
model for the origin of cosmic structure, particularly those
involving cosmic inflation$^{6}$, the initial fluctuations that
seeded the structure formation process form a Gaussian random
field$^{7}$. Deviations from uniformity, expressed in terms of the
density contrast $\delta({\bf x})$ defined by
\begin{equation} \delta ({\bf x}) = \frac{\rho({\bf
x})-\rho_0}{\rho_0}, \end{equation} where $\rho_0$ is the average
density and $\rho({\bf x})$ is the local matter density. Because
the initial perturbations evolve linearly, it is useful to expand
$\delta({\bf x})$  as a Fourier superposition of plane waves:
\begin{equation}
\delta ({\bf x}) =\sum \tilde{\delta}({\bf k}) \exp(i{\bf k}\cdot
{\bf x}).
\end{equation}
The Fourier transform $\tilde{\delta}({\bf k})$ is complex and
therefore possesses both amplitude $|\tilde{\delta} ({\bf k})|$
and phase $\phi_{\bf k}$ where
\begin{equation}
\tilde{\delta}({\bf k})=|\tilde{\delta} ({\bf
k})|\exp(i\phi_{\bf_k}). \label{eq:fourierex}
\end{equation}
Gaussian random fields possess Fourier modes whose real and
imaginary parts are independently distributed. In other words,
they have phase angles $\phi_k$ that are independently distributed
and uniformly random on the interval $[0,2\pi]$. When fluctuations
are small, i.e. during the linear regime, the Fourier modes evolve
independently and their phases remain random. In the later stages
of evolution, however, wave modes begin to couple together$^{4}$.
In this  regime the phases become non-random and the density field
becomes highly non--Gaussian. Phase coupling is therefore a key
consequence of nonlinear gravitational processes if the initial
conditions are Gaussian and a potentially powerful signature to
exploit in statistical tests of this class of models.

The information needed to fully specify a non-Gaussian field (or,
in a wider context, the information needed to define an
image$^{12}$) resides in the complete set of Fourier phases.
Unfortunately, relatively little is known about the behaviour of
Fourier phases in the nonlinear regime of gravitational
clustering$^{13-18}$, but it is of great importance to understand
phase correlations in order to design efficient statistical tools
for the analysis of clustering data. A vital first step on the
road to a useful quantitative description of phase information is
to represent it visually. We do this using colour, as shown in
Figure 2. To view the phase coupling in an N-body simulation one
Fourier-transforms the density field to produce a complex array
containing the real (R) and imaginary (I) parts of the transformed
`image' with the pixels in this array labelled by wavenumber ${\bf
k}$ rather than position ${\bf x}$. The phase for each wavenumber,
given by $\phi=\arctan(I/R)$, is then represented as a hue for
that pixel.

The rich pattern of phase information revealed by this method (see
Figure 3) can be quantified and related to the gravitational
dynamics of its origin. For example in our analysis of phase
coupling$^{5}$ we introduced a quantity $D_k$, defined by
\begin{equation}
D_k\equiv\phi_{k+1}-\phi_{k},
\end{equation}
which measures the difference in phase of modes with neighbouring
wavenumbers in one dimension. We refer to $D_k$ as the phase
gradient. To apply this idea to a two-dimensional simulation we
simply calculate gradients in the $x$ and $y$ directions
independently. Since the difference between two circular random
variables is itself a circular random variable, the distribution
of $D_k$ should initially be uniform. As the fluctuations evolve
waves begin to collapse, spawning higher-frequency modes in phase
with the original$^{22}$. These then interact with other waves to
produce the non-uniform distribution of $D_k$ seen in Figure 3.

It is necessary to develop quantitative measures of phase
information that can describe the structure displayed in the
colour representations. In the beginning the phases $\phi_k$ are
random and so are the $D_k$ obtained from them. This corresponds
to a state of minimal information, or in other words maximum
entropy. As information flows into the phases  the information
content must increase and the entropy decrease. This can be
quantified by defining an information entropy on the set of phase
gradients$^{5}$. One constructs a frequency distribution, $f(D)$
of the values of $D_k$ obtained from the whole map. The entropy is
then defined as
\begin{equation} S(D)=-\int f(D)\log [f(D)] dD,
\end{equation}
where the integral is taken over all values of $D$, i.e. from $0$
to $2\pi$. The use of $D$, rather than $\phi$ itself, to define
entropy is one way of accounting for the lack of translation
invariance of $\phi$, a problem that was missed in previous
attempts to quantify phase entropy$^{23}$. A uniform distribution
of $D$ is a state of maximum entropy (minimum information),
corresponding to Gaussian initial conditions (random phases). This
maximal value of $S_{\rm max}=\log(2\pi)$ is a characteristic of
Gaussian fields. As the system evolves it moves into to states of
greater information content (i.e. lower entropy). The scaling of
$S$ with clustering growth displays interesting properties$^{17}$,
establishing an important link between the spatial pattern and the
physics driving clustering growth. This phase information is a
unique fingerprint of gravitational instability and it therefore
also furnishes statistical tests of the presence of any initial
non-Gaussianity$^{24}$.

\newpage

\noindent{\bf Figure 1.} A numerical simulation of galaxy
clustering (left) together with a version of the same picture
generated by randomly reshuffling the phases between Fourier modes
of the original picture.  Since the amplitude of each Fourier mode
is unchanged in this operation, these two pictures have exactly
the same power-spectrum, $P(k)\propto|\tilde{\delta}({\bf k})|^2$,
but have totally different morphology. The shortcomings of $P(k)$
can be partly ameliorated by defining higher-order quantities such
as the bispectrum$^{4,8-10}$ or correlations$^{11}$ of
$\tilde{\delta}({\bf k})^2$. The bispectrum and higher-order
polyspectra vanish for Gaussian fields, but in a non-Gaussian
field they may be non-zero. The usefulness of these and related
quantities therefore lies in the fact that they encode some
information about non-linearity and non-Gaussianity. The
bispectrum, for example, measures the phase coupling induced by
quadratic nonlinearities, and so on for higher orders. However, an
infinite hierarchy of such moments is necessary to specify the
properties of a general random field in a statistical sense. Since
not every distribution is specified by its moments, even this need
not be complete.

\vspace{0.5cm}

\noindent{\bf Figure 2.} The representation of colour hue on a
circle.  In colour image display devices, each pixel represents
the intensity and colour at that position in the image$^{18,19}$.
The quantitative specification of colour involves three
coordinates describing the location of that pixel in an abstract
colour space, designed to reflect as accurately as possible the
eye's response to light of different wavelengths. In many devices
this colour space is defined in terms of the amount of Red, Green
or Blue required to construct the appropriate tone; hence the RGB
colour scheme. The scheme we are particularly interested in, the
HSB scheme, is based on three different parameters: Hue,
Saturation and Brightness. Hue is the term used to distinguish
between different basic colours (blue, yellow, red and so on).
Saturation refers to the purity of the colour, defined by how much
white is mixed with it. A saturated red hue would be a very bright
red, whereas a less saturated red would be pink. Brightness
indicates the overall intensity of the pixel on a grey scale. The
HSB colour model is particularly useful because of the properties
of the `hue' parameter, which is defined as a circular variable.
On the colour circle shown, the primary hues (red, green and blue)
are  $120\:^\circ$ apart from one another (at $0^\circ$,
$120^\circ$ and $240^\circ$ respectively), while the complementary
tones yellow, cyan, and magenta are at $60^\circ$, $180^\circ$ and
$300^\circ$. Red, of course also appears at $360^\circ$ and so on,
so this parameter is truly circular in the same way as phases are.
Our visualization method the hue parameter to encode the Fourier
phases.

\vspace{0.5cm}

\noindent{\bf Figure 3.} The evolution of phase coupling in a
sequence of snapshots from a two-dimensional N-body
simulation$^{20,21}$ starting with Gaussian initial conditions.
The initial power-spectrum was a power-law, $P(k)\propto k^n$,
with $n=0$. The left-hand column shows the development of
hierarchical clustering through the emergence in the density field
of the characteristic network of voids and filaments that typifies
the action of gravitational instability. The next column shows the
colour-coded phases, and the third and fourth columns show phase
gradients in the $x$-direction and $y$-direction respectively.
Each row corresponds to a particular timestep, with time flowing
downwards.  The early stages display no phase structure, but
patterns emerge as time unfolds. The phase images develop a series
of bands and stripes, first appearing as large-scale features in
reciprocal space. As the simulation evolves, the characteristic
$k$-space scale of these features gets smaller. This behaviour is
related to the increasing characteristic scale of features in the
density maps in the first column: larger features in real space
correspond to smaller features in the reciprocal space. Phase
correlations between $k$-modes reveal themselves as structure in
the distribution of the $D_k$, chiefly through a domination of
some hues over others. Starting from the initial data, in which
they are distributed evenly, one sees the gradual appearance of
regions where certain hues dominate. By the final stage, the third
column is largely green and the last column is largely red
indicating strongly coupled phases in both directions.
 Animations
of this effect can be viewed at our website
\begin{verbatim}
http://www.nottingham.ac.uk/~ppzpc/phases/index.html.
\end{verbatim}

Please address correspondence to Peter Coles
(Peter.Coles@Nottingham.ac.uk). Colour animations of phase
evolution from a set of $N$-body experiments, including the one
shown in Figure 3, can be viewed at
\begin{verbatim}
http://www.nottingham.ac.uk/~ppzpc/phases/index.html.
\end{verbatim}

\end{document}